# A model for processivity of molecular motors[*]

Ping Xie, Shuo-Xing Dou, and Peng-Ye Wang

*Laboratory of Soft Matter Physics, Institute of Physics, Chinese Academy of Sciences,*

*Beijing 100080, China*

ABSTRACT

We propose a two-dimensional model for a complete description of the dynamics of molecular motors, including both the processive movement along track filaments and the dissociation from the filaments. The theoretical results on the distributions of the run length and dwell time at a given ATP concentration, the dependences of mean run length, mean dwell time and mean velocity on ATP concentration and load are in good agreement with the previous experimental results.

Molecular motors are protein molecules that convert chemical energy to mechanical force. They are responsible for essentially all active biological motions [1]. An important class of molecular motors is transport proteins capable of moving processively and unidirectionally along linear, periodically structured and polar track filaments. They are divided into three superfamilies: kinesin, dynein, and myosin. These motor proteins perform such tasks as intracellular transport, cell division, bacterial motion, and muscle contraction [1-9].

Up to now, a comprehensive understanding of the microscopic mechanism of molecular motors is still challenging. But with the extensive investigations using different experimental methods, such as biochemical, biophysical, and single-molecular approaches, many aspects of the movement behavior of different molecular motors have been gradually elucidated, and a large amount of data been gathered. In the case of kinesin, for example, the rapid development and progress of single-molecule manipulation and detection techniques in recent years have improved significantly our knowledge of its dynamic and mechanistic properties *in vitro* [10-20]. In particular, in Ref. [18-20] detailed single-molecular observations of the mechanical behavior of kinesin moving along microtubules *in vitro* under low load or under controlled loads have been reported. Important parameters such as stall force, velocity, mean run length, and dwell time have been measured systematically.

In proportion with experimental studies, molecular motors have also been extensively studied theoretically. Theoretical modeling of the motion of motor proteins involves mainly two approaches. One of the approaches uses the traditional chemical kinetic descriptions [19,21,22]. The other approach is based on thermal ratchet models in which the molecular motor is treated as a Brownian particle [23-29]. The processivity of molecular motors such as the average attachment time to the filament and the mean run length has been studied in the frame of one-dimensional two-state model [24]. With periodic sequential kinetic models the probability of detachment of the molecular motors from the filament has been considered [30].

---

[*] Project supported by the National Natural Science Foundation of China (Grant No 60025516).

The aim of this Letter is to provide a simple two-dimensional model to describe the dynamics of molecular motors and give comparisons of our theoretical results with previous experimental results. In order to study both the processive movement of the molecular motor along a periodic and polar track filament and its dissociation from the filament, we consider an overdamped Brownian particle in two spatial dimensions, with $x$-axis along the filament and $y$-axis perpendicular to the filament. We use the following Langevin equations:

$$\frac{dx}{dt} = -\frac{\partial U(x,y)}{\partial x} + \sqrt{2D}\xi_x(t), \qquad (1a)$$

$$\frac{dy}{dt} = -\frac{\partial U(x,y)}{\partial y} + \sqrt{2D}\xi_y(t), \qquad (1b)$$

where, for convenience, the viscosity is scaled away. $\xi_x(t)$ and $\xi_y(t)$ are the stochastic Langevin noises representing the fluctuating forces: $\langle\xi_i(t)\rangle = 0$, $\langle\xi_i(t)\xi_j(t')\rangle = \delta_{ij}\delta(t-t')$, with $i, j$ representing $x$ or $y$. And $D$ is the noise strength.

In Eq. (1), $U(x, y)$ is the potential which is assumed to take the form

$$U(x,y) = u_1(y)[1-\cos(\pi x)] + u_2(x)[e^{-\frac{2(y-y_0)}{a}} - 2e^{-\frac{y-y_0}{a}}] - \alpha_x x - \alpha_y y. \qquad (2)$$

Here term $1-\cos(\pi x)$ represents the potential resulted from the periodic structure of the track filament with a period of 2 in the dimensionless Eqs. (1) and (2). In the case of kinesin, this period corresponds to 8 nm of microtubule filaments [10]. $u_1(y)$ in the first term of Eq. (2) is assumed to take the form $u_1(y) = u_{10}e^{-\frac{y-y_0}{b}}$, which means that, as the motor departs from the track filament, the potential resulted from the filament becomes weaker. Furthermore, we assume that the interaction between molecules of the motor and the filament is via a van der Waals force. Thus we adopt the Morse potential [31] as expressed by the second term in Eq. (2), where $u_2(x)$ is generally a periodic function of $x$, with the same period as that of the track filament.

The effect of energy released by the hydrolysis of ATP to ADP and $P_i$ is phenomenologically modeled as an effective force $\boldsymbol{\alpha}$ acting on the molecular motor, which is decomposed into component $\alpha_x$ along the track filament and component $\alpha_y$ perpendicular to the filament. This is described by the last term, $-\alpha_x x - \alpha_y y$, in Eq. (2). To obtain the dependence of the magnitude of $\boldsymbol{\alpha}$ on ATP concentration [ATP], we resort to the experimental results on stall force $F_{stall}$ in Ref.[18]. The measured $F_{stall}$ is the force necessary to stop the motion of the kinesin motor and, therefore, it is equivalent to that acting on the kinesin motor along the $x$ axis, $\alpha_x$, that is generated by the hydrolysis of ATP. A good fit to the measured $F_{stall}$ is shown in Fig. 1, where the fitting curve has the form $F_{stall} = c_1[1 - e^{-([ATP]/c_2)^{c_3}}]$ with $c_1$=7.6pN, $c_2$=3.0μM, and $c_3$=0.39. Note that, for a given system, the direction of $\alpha$ from ATP hydrolysis is fixed, i.e., the ratio between the two components $\alpha_x$ and $\alpha_y$ should be kept constant, and therefore the [ATP] dependence of both $\alpha_x$ and $\alpha_y$ can be expressed as $\alpha_i = g_i[1 - e^{-([ATP]/3.0)^{0.39}}]$ ($i = x, y$), where $g_x$ and $g_y$ represent the components along $x$ and $y$ directions, respectively, of the force generated by ATP hydrolysis at saturating [ATP]. When the motor is far away from the filament, the ATP-driving force should become zero and the motor is diffused freely in the noised environment according to Einstein's law. For clarity, the two-dimensional potential is schematically illustrated in Fig. 2.

In this Letter we numerically solve Eqs. (1) and (2). For simplicity, we neglect the dependence of the perpendicular van der Vaals force on the $x$ coordinate, i.e., assuming $u_2(x) = u_{20}$. Throughout the work we use $u_{10} = 1$ and $u_{20} = 3.8$, leading to comparable potential barrier heights in the two spatial dimensions. The equilibrium distance of the Morse potential is taken to be $y_0 = 0.1$ which is an order of magnitude smaller than the period of the filament. The breadth of the attractive region is $a = \frac{1}{2}y_0$ [32] and the decay length is $b = 2.0$ which is comparable to the period of the filament. As we have verified numerically, the variation of $y_0$, $a$ and $b$ has no essential effect on dynamical behaviors of the molecular motors. In Eqs. (1) and (2), only variations in the relative values of the five parameters, $u_{10}$, $u_{20}$, $g_x$, $g_y$, and $\sqrt{D}$, change the theoretical results, and a proportional variation in their absolute values can be scaled to time $t$ and thus makes no difference to the results.

Figure 3 shows one typical result of the time evolution of the motor displacements along $x$ and $y$. It is clear that the motor moves processively along the track filament in discrete steps of 2. The motor dissociates from the filament at $t = 2632.6$ and then is in free diffusion. This result shows striking resemblance to the experimental result in the literature [10,13,15].

To see the processive motility of the motor, we plot the distributions of the run length (the distance that the motor travels before dissociating and diffusing away from the filament) and the dwell time (the duration of each such run) for two ATP concentrations, which are shown in Fig. 4. Here and in all the following calculations we scale the distance $\Delta x = 1$ to 4 nm and the time $\Delta t = 1$ to 6.7 ms, in order to directly compare with the experiment for kinesin. It is seen that both the run length and the dwell time are exponentially distributed at any particular [ATP], which is consistent with the experimental results [20]. For further comparison with experiments in Ref. [20], we calculate the mean run length $L_m$, the mean velocity $V_m$, and the mean dwell time $T_m$, as a function of [ATP]. The mean values are obtained from 100 different realizations. In Fig. 5(a) we plot $L_m$, $V_m$, and $T_m$ versus [ATP] for $g_x = 2.8$ and $g_y = 6$, where data for velocity are fitted to Michaelis-Menten curve, $V = V_{\max}[\text{ATP}]/([\text{ATP}] + K_m)$. It is noted that our results show good agreement with the experimental results. In particular, the mean run length shows very weak dependence on [ATP] at large [ATP] and decreases a bit at low [ATP]. In Fig. 5(b) we show the mean run length $L_m$ for different sets of $g_x$ and $g_y$. It is seen that, for fixed $g_y = 6$, $L_m$ versus [ATP] shows the same tendency when varying $g_x$, except that its values increases with the increase of $g_x$. For $g_x = 2.8$ and $g_y = 5$, we see that the mean run length $L_m$ decreases obviously at low [ATP]. This is consistent with the report of Schnitzer *et al*. at low load [19].

To see the influence of the load on the processive motility of the motor, we add a negative force on the right-hand side of Eq. (1a). This means that a load is exerted along the track filament in the direction opposite to the motor movement, which is consistent with the case of experiment [19]. In Fig. 6(a) we plot $L_m$ versus load for two different [ATP]. It is seen that our results show good agreement with the experimental results [19]. Similar to the case of no load, we verified that, for a given load, the mean velocity also follows well the Michaelis-Menten dependence on [ATP], with some typical results shown in Fig. 6(b). This is consistent with the experimental results of Ref. [18].

Finally, we address the effect of the fluctuating force on the processive motility of the molecular motor. To this end, we calculate the dependences of the mean run length and the mean velocity on the noise strength $D$, which are shown in Fig. 7. It is seen that the mean velocity increases nearly linearly with the increase of $D$ in the range of the calculations. From this point of view, the larger the fluctuating force the more efficient the motor would be. However, the results also show that the mean run length decreases drastically with the increase of $D$. Thus although there is no experimental result for comparison in this aspect, we can imagine that, *in vivo*, $D$ should fall in a proper (not very wide) range to give both a large moving velocity and an adequate mean run length.

In summary, we propose a two-dimensional model to describe the dynamics of molecular motors, including both processive movement and dissociation. Even though this model is simple, it can give essential features of the motility of molecular motors. The good agreement between theoretical and previous experimental results for kinesin implies that this model represents a tenable theoretical approach to molecular motors.

**FIGURES**

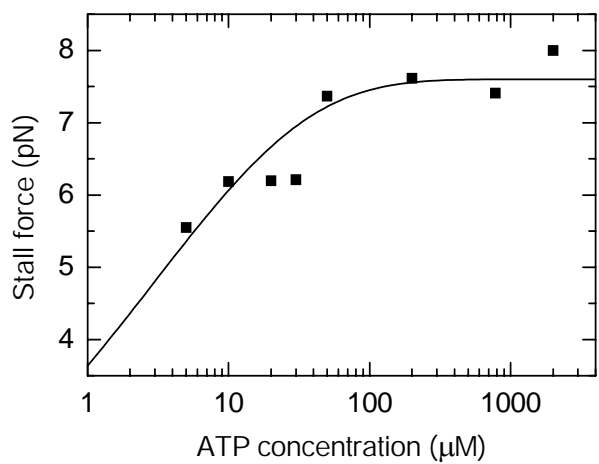

FIG. 1. Stall force versus [ATP]. Square dots are experimental data from Fig. 3b of Ref. [18] and the solid curve is a fit of the data to $F_{stall} = c_1[1-e^{-([ATP]/c_2)^{c_3}}]$ with $c_1$=7.6pN, $c_2$=3.0μM, and $c_3$=0.39.

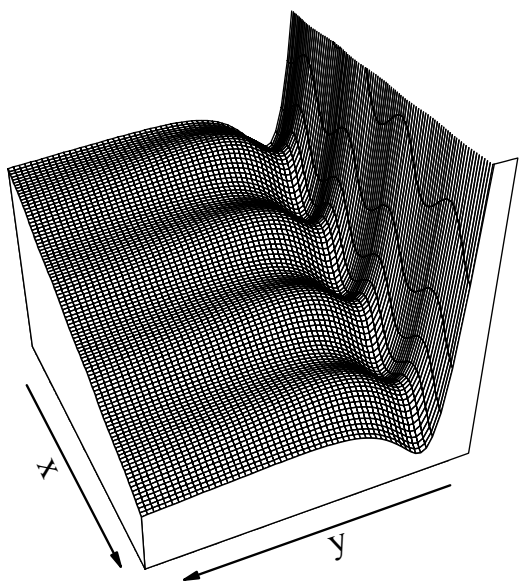

FIG. 2. A schematic plot of the two-dimensional potential $U(x, y)$ given by Eq. (2).

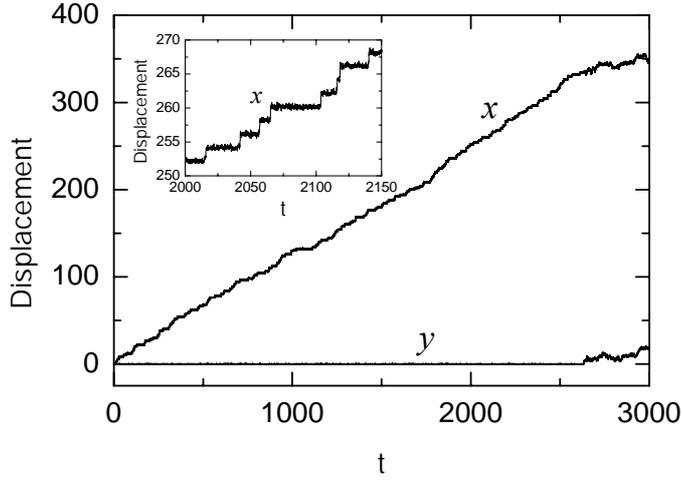

FIG. 3. Temporal evolution of the displacements of the molecular motor along $x$ and $y$ at [ATP] = 3 μM, $g_x$ = 2.8, $g_y$ = 6, and $D$ = 0.28. When $y \geq 10 y_0$ we take $g_x = g_y = 0$. Inset is a local enlargement.

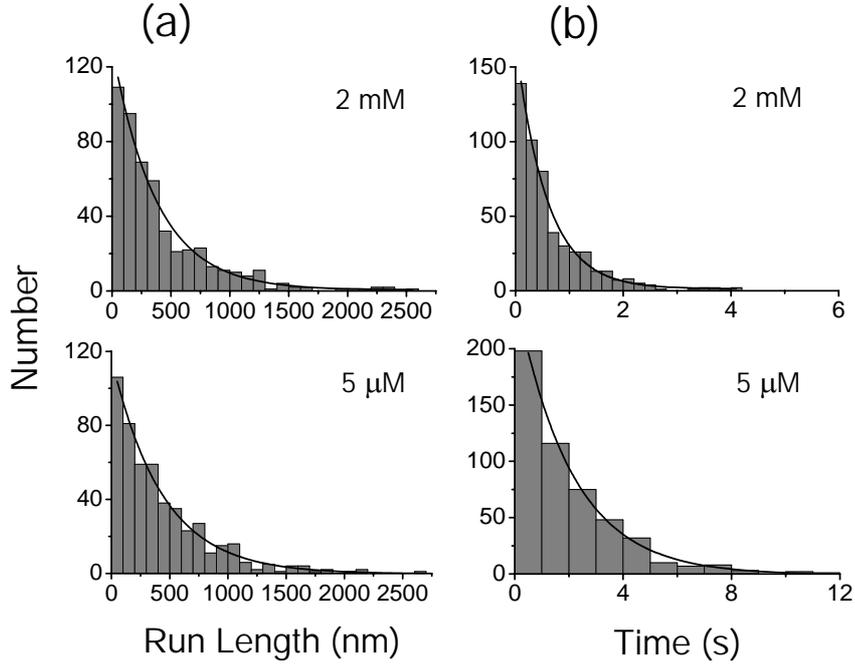

FIG. 4. Distribution of run length (a) and dwell time (b) at [ATP] = 2 mM (upper) and [ATP] = 5 μM (lower). $g_x$ = 2.8, $g_y$ = 6, and $D$ = 0.28.

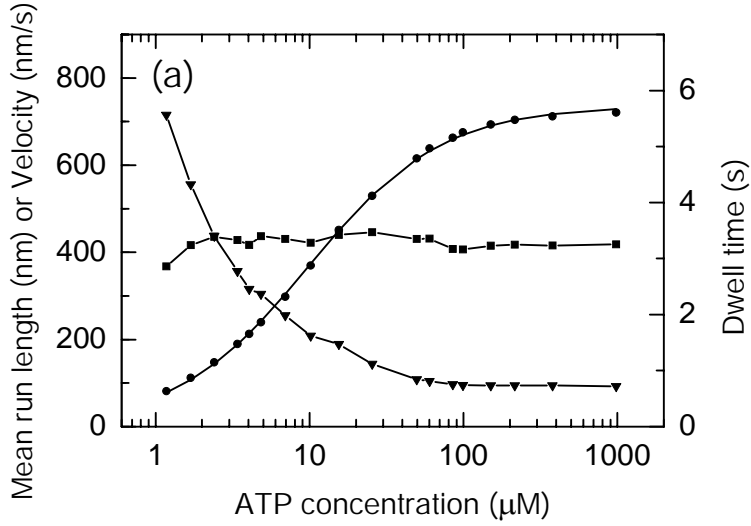

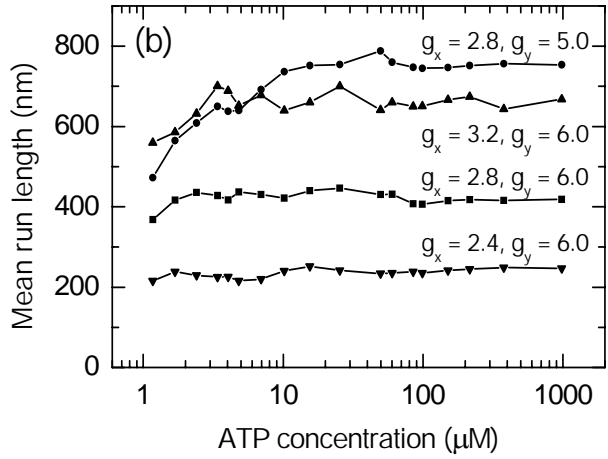

FIG. 5. (a) Mean run length (square dots), mean velocity (circular dots), and mean dwell time (triangular dots) versus [ATP] at $g_x$ = 2.8, $g_y$ = 6. Data for mean velocity are fitted to Michaelis-Menten curve, $V = V_{max}[\text{ATP}]/([\text{ATP}] + K_m)$, with $V_{max} = 735.95$ nm/s and $K_m = 9.84\mu M$. (b) Mean run length versus [ATP] at different $g_x$ and $g_y$. $D$ = 0.28.

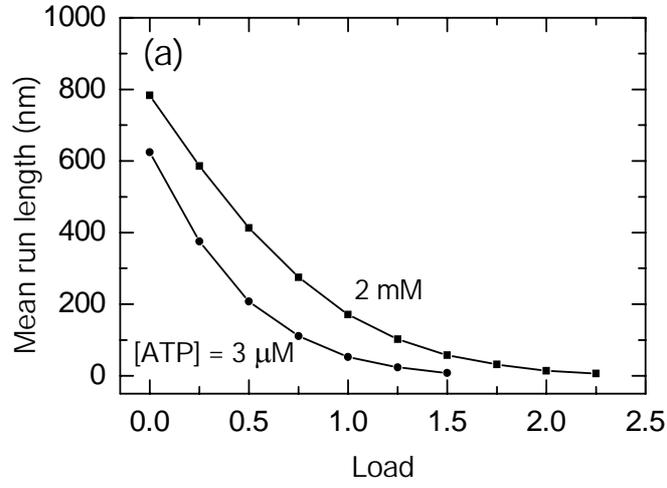

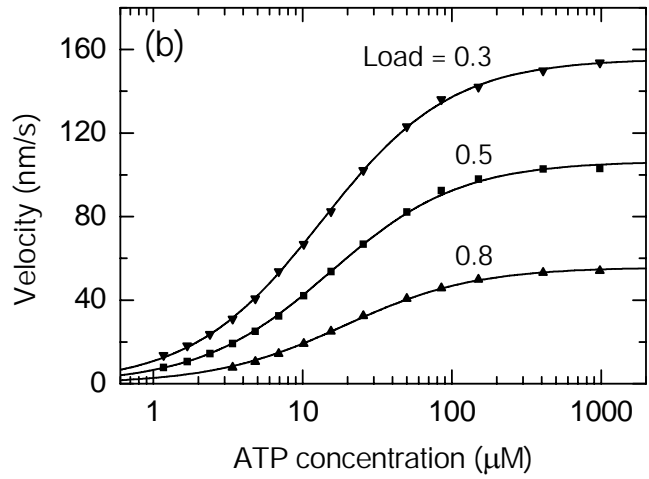

FIG. 6. (a) Mean run length versus load (in dimensionless units) at different [ATP] for $g_x = 2.8$, $g_y = 5$, and $D = 0.28$. (b) Mean velocity versus [ATP] at different loads for $g_x = 2$, $g_y = 4$, and $D = 0.28$. Data are fitted to Michaelis-Menten curves. $V_{max} = 155.59$ nm/s, $K_m = 13.38 \mu M$, Load = 0.3; $V_{max} = 106.52$ nm/s, $K_m = 15.22 \mu M$, Load = 0.5; $V_{max} = 55.76$ nm/s, $K_m = 19.12 \mu M$, Load = 0.8.

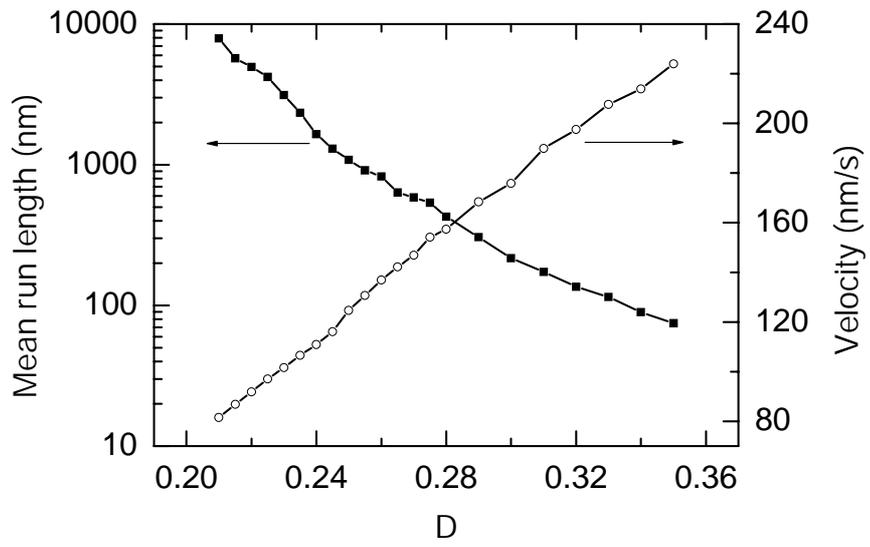

FIG. 7. Mean run length and mean velocity versus noise intensity $D$ (in dimensionless units). [ATP] = 3 μM, $g_x$ = 2.8, $g_y$ = 6.